\documentstyle[12pt,psfig,html]{article}
\def\reference{\parskip 0pt\par\noindent\hangindent 0.5 truecm}

\def\deg{\ifmmode ^{\circ}\else $^{\circ}$\fi}
\def\lapp{\ifmmode\stackrel{<}{_{\sim}}\else$\stackrel{<}{_{\sim}}$\fi} 
\def\gapp{\ifmmode\stackrel{>}{_{\sim}}\else$\stackrel{>}{_{\sim}}$\fi} 
\def\farcs{\hbox{$.\!\!^{\prime\prime}$}}
\begin{document}
%
%
\title{Long-term Monitoring of Molonglo Calibrators}
%


\author{B. M. Gaensler $^{1,2}$\thanks{Hubble Fellow} \and
 R. W. Hunstead $^{2}$ \and
} 

\date{}
\maketitle

{\center
$^1$ Center for Space Research, Massachusetts Institute of Technology,
70 Vassar Street, \\ Cambridge MA 02139, USA; bmg@space.mit.edu \\[3mm]
$^2$ School of Physics, University of
Sydney, NSW 2006, Australia; rwh@physics.usyd.edu.au \\[3mm]
}

%
\begin{abstract}

Before and after every 12 hour synthesis observation, 
the Molonglo Observatory Synthesis Telescope (MOST)
measures the flux densities of $\sim$5  compact
extragalactic radio sources, chosen from a list of 55 calibrators. 
From 1984 to 1996, the MOST made some 58\,000 such measurements.
We have developed an algorithm to process this dataset to produce
a light curve for each source spanning this thirteen year period.
We find that 18 of the 55 calibrators are variable,
on time scales between one and ten years. There is the tendency
for sources closer to the Galactic Plane to be more likely to vary,
which suggests that the variability is a result
of refractive scintillation in the Galactic interstellar medium.
The sources with the flattest radio spectra show the highest 
levels of variability,
an effect possibly resulting from differing orientations of the
radio axes to the line of sight.

\end{abstract}


\bigskip

%
%

\section{Introduction}


Many compact extragalactic radio sources show variations in their radio
flux density as a function of time. At high frequencies ($\nu \gg
1$~GHz) this variability is usually interpreted as being
intrinsic to the source (e.g.\ Qian et al 1995, although
see Kedziora-Chudczer et al 1997).
%
Variability at lower frequencies (e.g.\ Hunstead 1972;
Ghosh \& Rao 1992) is normally attributed to
refractive interstellar scintillation, in which the
intensity variations are caused by distortions of the wavefront by
electron density gradients in an intervening screen of material
(Shapirovskaya 1978; Rickett et al 1984).  There is evidence that
the parameters of such variability depend on the 
Galactic latitude of the source (Spangler et
al. 1989; Ghosh \& Rao 1992), suggesting that the material causing the
scintillation is in our own Galaxy.

In some sources, both intrinsic variability and scintillation may be
occurring at the same time (e.g. Mitchell et al 1994). Such sources
show large but uncorrelated variations at  high and low
frequencies. At frequencies $\nu \approx 1$~GHz, one might expect both
effects to occur; however, variability in this region of the spectrum
is largely unexplored. The Molonglo Observatory Synthesis Telescope
(MOST; Mills 1981; Robertson 1991) operates at a frequency of
843~MHz, and is thus well placed to study this regime. For calibration
purposes, the MOST monitors the flux density of $\sim$10 compact
extragalactic sources every day. Thus the full record of MOST
calibrations, running from 1984 until the commencement of the Wide
Field Project in 1996 (Large et al 1994), 
forms an ideal database with which to study
variability in this intermediate frequency range.

A preliminary analysis of three MOST calibrators was made by
Campbell-Wilson \& Hunstead (1994), hereafter Paper I.  It was shown
that flux density measurements with a relative accuracy of 2\% could be
extracted from the database. Over the period from 1990.1 to 1993.7, the
source MRC~B0409--752 was shown to be stable, while MRC~B0537--441 and
MRC~B1921--293 were found to be highly variable. In this paper we now
report on all 55 calibrators used by the MOST, over a thirteen year period.
In Section~\ref{obs} we explain how we process the calibrator measurements
in order to produce light curves for each source, and then determine
whether a source is variable or not. In Section~\ref{results} we present
light curves for all 55 sources, plus structure functions for those
sources found to be variable. In Section~\ref{discuss} we discuss
some individual sources in our sample, and consider whether any of
the observed properties correlate with Galactic latitude.

\section{Observations and Data Analysis}
\label{obs}

\subsection{SCAN Measurements}
\label{obs_scan}

The MOST is an east-west synthesis telescope, consisting
of two cylindrical paraboloids of dimensions 778~m~$\times$~12~m.
Radio waves are received by a line feed system of 7744 circular
dipoles.
The telescope is steered by mechanical rotation of
the cylindrical paraboloids about their long axis,
and by phasing the feed elements along the arms.
In a single 12-hour synthesis, the MOST can produce
an image at a spatial resolution of $43''\times43''{\rm cosec}(|\delta|)$
and at a sensitivity of $\sim$1~mJy~beam$^{-1}$ (where
1 jansky~[Jy]~$=10^{-26}$~W~m$^{-2}$~Hz$^{-1}$).

Before and after each 12-hour synthesis, the MOST typically observes
$\sim5$ calibration sources in fan-beam 
``SCAN'' mode in order to determine the gain
and pointing corrections for the telescope. These sources are
chosen from a list of 55 calibrators, 45 of which
were chosen from the Molonglo Reference Catalogue (MRC)
at 408~MHz (Large et al 1981), using as selection criteria that they have
declination $\delta < -30^{\circ}$, Galactic latitude
$|b| > 10^{\circ}$, angular sizes $<10''$ and flux densities 
$S_{\rm{408\,MHz}}>4$~Jy
and $S_{\rm{843\,MHz}}>2.5$~Jy; further discussion is
given by Hunstead (1991). This list was later
supplemented by seven flat-spectrum ($S_{\rm{408\,MHz}}<4$~Jy) sources
from the work of Tzioumis (1987), plus three compact sources
for which $\delta > -30^{\circ}$. The full list of calibrators is
given in Paper I.

\begin{figure}
\centerline{\psfig{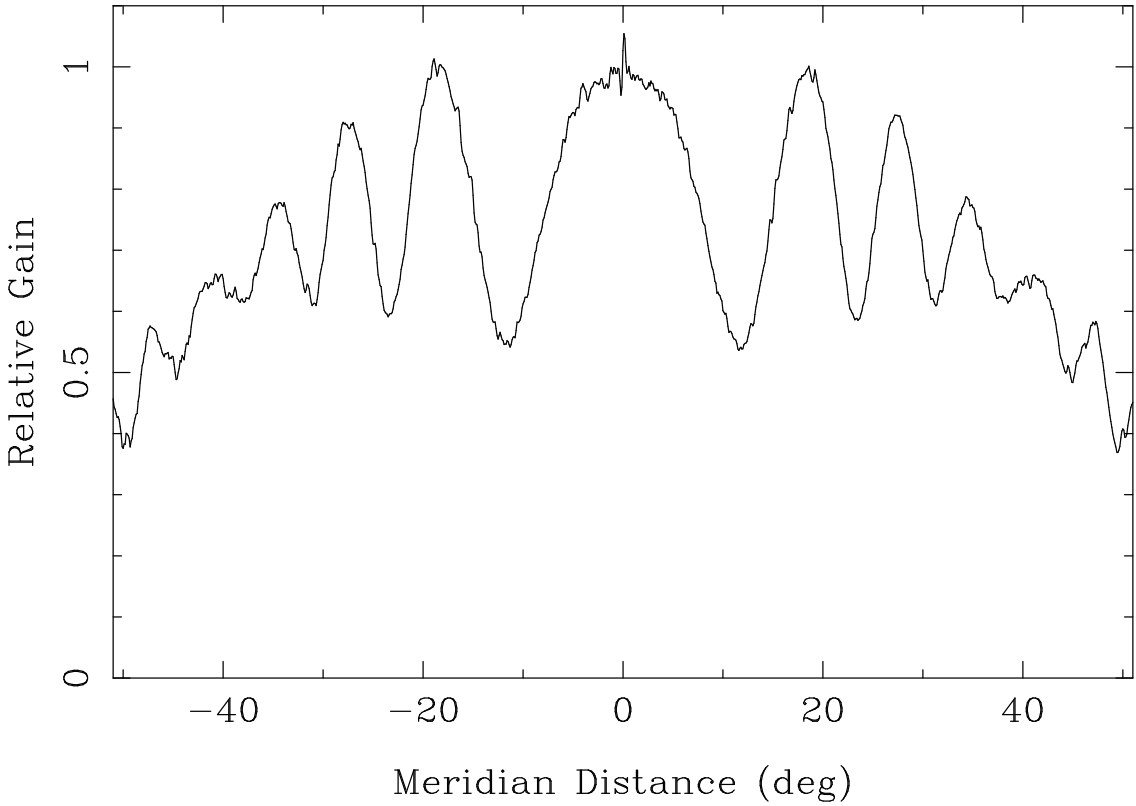}}
\caption{The relative gain of the MOST as a function of meridian
distance. The data shown correspond to the asymmetric gain curve
of Burgess (priv comm), with additional
empirical corrections of order 2\%.}
\label{fig_md}
\end{figure}

For each SCAN observation the calibrator source is tracked for two
minutes, after which the mean antenna response is compared with the
theoretical fan-beam response to a point source. 
From 1994 to 1996, over 58\,000 such measurements were made.
In each case, parameters such as the goodness-of-fit of the response
and the pointing offset from the calibrator position are
recorded, along which an amplitude which is the product of the
instantaneous values of the source flux density, the intrinsic
telescope gain and local sensitivity factors. The main
factors are strong but well-determined functions of
meridian distance\footnote{Meridian
distance, MD, is related to hour-angle, $H$, by $\sin {\rm MD} \approx
\cos \delta \sin H$ --- see Robertson (1991).} (MD) and of ambient
temperature (which ranges from $-$10\deg C to +40\deg C during the
year); the variation of sensitivity with MD is shown in
Figure~\ref{fig_md}. After applying corrections for these two
factors, the telescope gain for each SCAN is derived by
comparing the corrected amplitude with the tabulated flux
density of the corresponding source (see Table~1 of Paper~I).
The residual scatter in the gain determined from steady sources
(defined in Section~\ref{obs_analysis}) is typically 2\% RMS; this is
the fundamental limit to the uncertainty of measurements made
using the SCAN database.

\subsection{Selection Criteria}
\label{obs_select}

Various selection criteria are applied to the SCAN database before
accepting measurements for further analysis:

\begin{itemize}

\item{} The uncertainties in the 
MD gain curve increase towards
large MD, and observations made outside the MD range
$\pm50$\deg\ are excluded;

\item{} Observations made during routine performance testing (characterised by
a large number of successive SCANs of the same source) are discounted,
except where the standard deviation in gain was less than 5\%. In such
cases the group is treated as a single measurement with a gain equal to the
average of the group;

\item{} a poor fit to the antenna response can
often indicate a confusing source or a telescope malfunction, and
such data are excluded;
\item{} extreme values of the relative gain (below 0.5 or above 1.5) are
assumed to be discrepant and are discarded.

\end{itemize}


Because calibration observations are made just before and after each
synthesis, the database is typically clustered into SCANs
closely spaced in time. We define a ``block'' as a group of at least
three valid observations made within the space of an hour. We initially
exclude observations of 15 of the 55 calibrators (see Table~1 of Paper~I),
because of: (i) a flat spectrum ($\alpha > -0.5$, $S \propto
\nu^{\alpha}$), (ii) suspected variability or (iii)
the presence of a confusing
source.  By averaging the gains determined from each SCAN within a
block, a representative gain for the telescope at that particular epoch
can be determined. This is then applied to each individual
observation within the block to obtain a measurement of 
flux density for that source.

Some of the resultant light curves have thousands of data points,
generally sampled at highly irregular intervals. Some light curves have
significant scatter; it is not clear whether this scatter is due to
unrecognised systematic errors in our flux density determination,
to true variability on time-scales shorter than the
typical sampling interval, or to the presence of confusing sources
in the field. In any case, we chose to bin each light-curve at
30 day intervals; the mean of all flux densities within a given bin
becomes a single point on a smoothed light curve, and the standard
deviation of the measurements in that bin becomes the error bar
associated with this measurement\footnote{In cases where there
is only one measurement in a particular 30 day interval, the error is
nominally assigned to be 5\% of the measured flux density.}. While
binning the data filters out any genuine variability on time-scales
less than a month, the irregular sampling intervals of the observations
and the inherent uncertainty in a single SCAN's flux density make the
MOST database less than ideal for studying such short-term behaviour.

\subsection{Analysis of Variability}
\label{obs_analysis}

In order to quantify
which sources are variable and which are steady,
we calculate the $\chi^2$ probability
that the flux has remained constant for a given source
(e.g.\ Kesteven et al 1976). We first calculate the quantity
\begin{equation}
x^2 = \sum_{i=1}^{n} (S_i - \tilde{S})^2/\sigma_i^2
\end{equation}
where $\tilde{S}$ is the weighted mean, given by
\begin{equation}
\tilde{S} = \frac{\sum_{i=1}^{n} (S_i/\sigma_i^2)}{\sum_{i=1}^{n} 
(1/\sigma_i^2)},
\end{equation}
$S_i$ is the $i$th measurement of the flux density for a particular
source, $\sigma_i^2$ is the variance associated with each
30-day estimate of $S_i$, and $n$ is 
the number of binned data points for that source. 
For normally-distributed random errors, we expect $x^2$ to be 
distributed as $\chi^2$ with $n-1$ degrees of freedom. For each
source, we can then calculate the probability, $P$, of exceeding $x^2$ by chance
for a random distribution. 

A high value of $P$ indicates that a source has a steady flux density
over the available time period; we classify a source as 
steady (S) if $P>0.01$, and undetermined (U) if $0.001 < P < 0.01$. 
However the $\chi^2$ test cannot distinguish between
sources which are genuinely variable
and those which simply have a large scatter in their light curve; both
light curves result in a low value of $P$. We distinguish between
these possibilities by computing the structure function (e.g. Hughes et al 
1992; Kaspi \& Stinebring 1992) for each
source for which $P<0.001$. 
The mean is subtracted from the binned time series $S_t$,
and these data are then normalised
by dividing by the standard deviation. This yields a 
new time series $F_t$, from which the structure function
\begin{equation}
\Sigma_\tau = \langle [ F_{t + \tau} - F_t]^2 \rangle
\end{equation}
can be calculated, 
where $\tau$ is a parameter known as the lag. 
If a light curve contains scatter but no true variability, then
the structure function will have the value
$\Sigma_\tau \approx 2$ for all values of $\tau$. But when a source is truly varying,
we expect the resulting structure function to consist of three regimes:

\begin{itemize}

\item{Noise regime:}
at small lags, $\Sigma_\tau$ is more or less constant.

\item{Structure regime:}
as $\tau$ increases, $\Sigma_\tau$
increases linearly (on a log-log plot). 

\item{Saturation regime:}
at high lags, the structure function 
turns over and oscillates around $\Sigma_\tau = 2$ (for our
normalisation). If there is a second, longer, time-scale
in the data, the structure function can enter another
linear regime at longer lags before again saturating.
\end{itemize}

If a source has $P<0.001$ but shows no clear structure in its structure
function, we classify it as undetermined (U).  Only sources which 
have both $P < 0.001$ and show structure are classified as variable (V). In
these cases, the structure function can also be used to obtain a
characteristic time scale, $\tau_V$, for variability; we define
$\tau_V$ to be equal to twice the lag at which the structure function
saturates. We expect a structure function to be sensitive only
to time scales longer than about 100 days (i.e. a few multiples of
the sampling interval of the binned data). Furthermore, caution
should be applied when interpreting structure at
large values of $\tau$, as only a few points make a contribution
to $\Sigma_\tau$ at these long lags (e.g. Hughes et al 1992).


\section{Results}
\label{results}

\begin{figure}
\centerline{\psfig{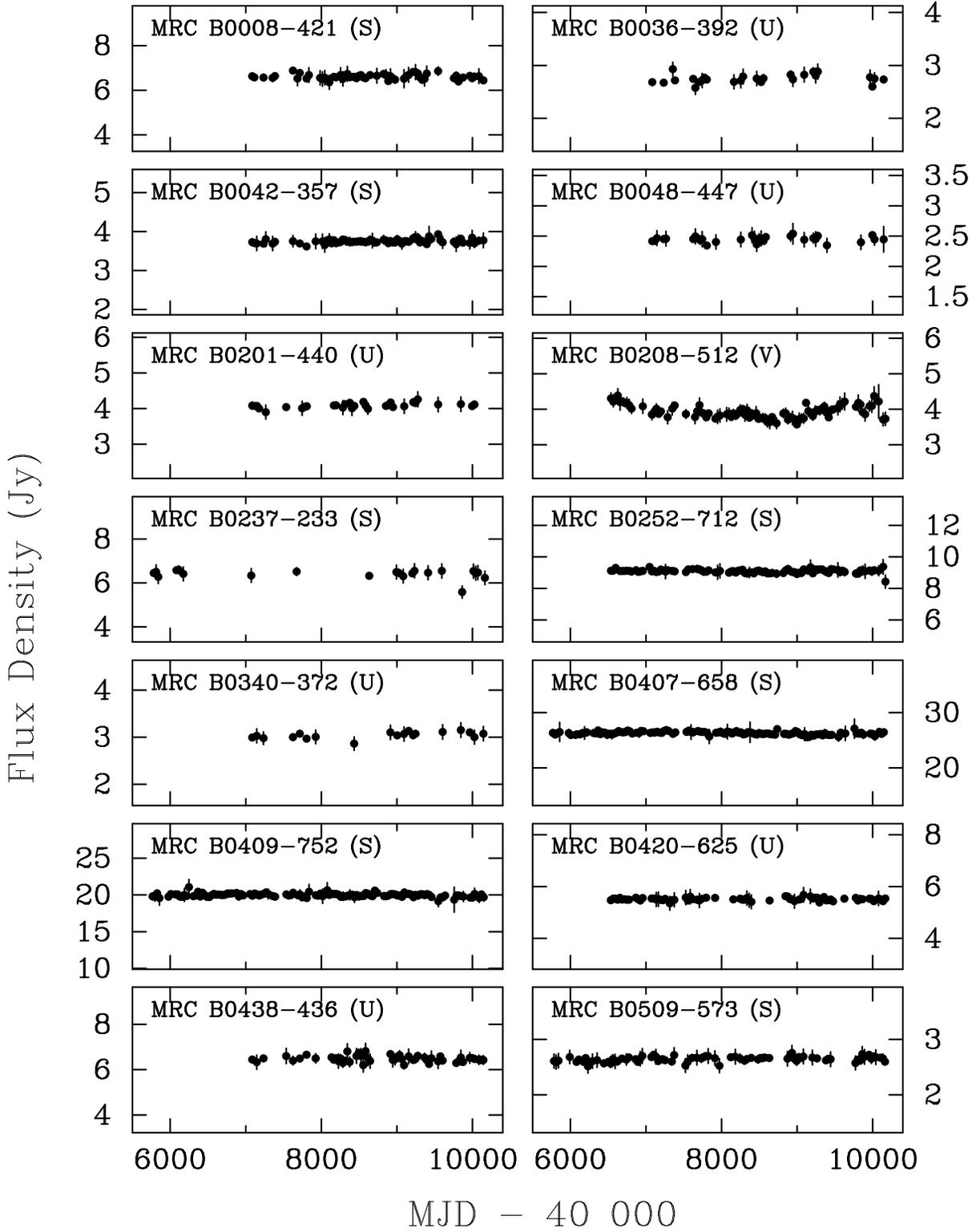}}
\caption{Light curves for 55 MOST calibrators. Sources
are marked S, U or V, corresponding to whether their
time behaviour is steady, undetermined or variable
respectively. Each
abcissa ranges between
MJD -- 40\,000 $=$~5500 (1983 Jun) and 10400 (1996 Nov), while 
ordinates run between
0.5 and 1.5 times the nominal flux density for each source (see Table~1
of Paper~I). Data have been binned into 30 day intervals --- the error bars
shown are the 1$\sigma$ standard deviation of the individual data points within
each interval, or are set at 5\% in cases where only one data point falls
in a given 30 day period.}
\label{fig_sources_1}
\end{figure}

Approximately 28\,000 SCANs meet the selection criteria described in
Section~\ref{obs_select}, and around 22\,000 of these fall within valid
blocks. The resulting light curves for the 55 MOST calibrators are given
in Figure~\ref{fig_sources_1}.\footnote{The corresponding data tables
are available at http://www.physics.usyd.edu.au/astrop/scan/ .}
Using the criteria described in
Section~\ref{obs_analysis}, 18 sources are found to be variable, 19 are
found to have steady light curves, and the remaining 18 are
undetermined. Each source in Figure~\ref{fig_sources_1} is marked with
a V, S or U corresponding to its classification.

\begin{figure}
\centerline{\psfig{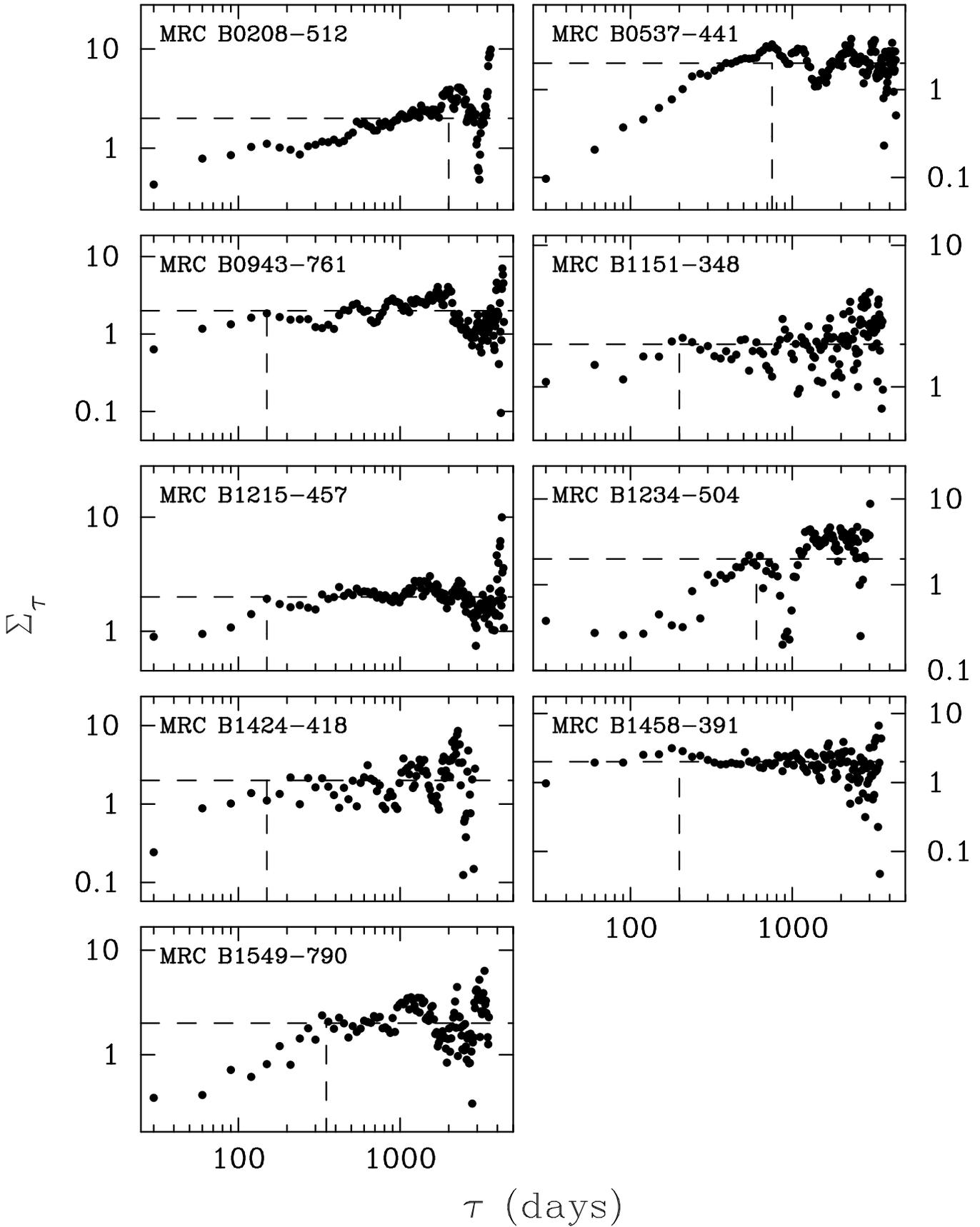}}
\caption{Structure functions for the 18 variable sources.
The dashed horizontal line corresponds to $\Sigma_\tau = 2$,
the value at which a pure sinusoid will saturate. The dashed
vertical line corresponds to the approximate lag at which
the structure function saturates (the time scale for variability,
$\tau_V$, is defined to be twice this value).}
\label{fig_struc}
\end{figure}

Structure functions for the 18 variable sources are shown in
Figure~\ref{fig_struc}; for each source we have estimated the time
scale for variability, $\tau_V$, as marked on each plot. However, we
note that some of these estimates are very approximate, as a result of
the sparse and/or irregular sampling of the light curves. For example,
for MRC~B2326--477 we have assigned $\tau_V = 400$~d, but one could
just as easily argue that $\tau_V = 2000$~d. Furthermore, there is
evidence that the structure functions for some sources, such as
MRC~B1740--517, enter another
linear regime beyond the point where they saturate. This suggests
that there are variations on time scales
longer than we can measure with these data.
Some properties of the 18 variable sources are summarised
in Table~\ref{tab_variables}.

\section{Discussion}
\label{discuss}

\subsection{Individual Sources}

We restrict our comments here to the 18 sources found to be variable.
Many of these sources have been observed in snapshot mode at 5 GHz
with the Australia Telescope Compact Array (ATCA, Burgess 1998), and
are also ATCA secondary phase calibrators.  

\medskip
\noindent{\bf MRC B0208$-$512}: VLBI modelling shows a strong core
(Preston et al 1989), and a jet-like feature (Tingay et al 1996).
Detected as an X-ray source in the {\em ROSAT}\ All-Sky Survey (Brinkmann et
al.\ 1994) and as a $\gamma$-ray source in the {\em EGRET}\ survey (Bertsch et
al.\ 1993).

\noindent{\bf MRC B0537$-$441}: See Paper I.

\noindent{\bf MRC B0943$-$761}: Close $2\farcs8$ double at 5 GHz (Burgess
1998).  Detected as an X-ray source in the {\em ROSAT}\ All-Sky Survey
(Brinkmann et al 1994).

\noindent{\bf MRC B1151$-$348}: Radio spectrum peaks at $\sim$200 MHz.
A VLBI image shows a 90~mas double structure (King et al 1993).

\noindent{\bf MRC B1215$-$457}: Compact steep-spectrum source with a
strong, slightly resolved VLBI core (Preston et al 1989).

\noindent{\bf MRC B1234$-$504}: Compact steep-spectrum source, with no
optical counterpart on the UK Schmidt sky survey but possible stellar
identification on a CCD image obtained at the Anglo-Australian
Telescope (AAT) (Burgess 1998).

\noindent{\bf MRC B1424$-$418}: Discordant flux densities measured at
Parkes point to the source being variable at~5 GHz (Burgess, priv
comm).  VLBI modelling shows an unequal 23 mas double structure
(Preston et al 1989).

\noindent{\bf MRC B1458$-$391}: Compact steep-spectrum source in a
crowded optical field; optical ID based on an AAT CCD image (Burgess
1998).

\noindent{\bf MRC B1549$-$790}: VLBI image shows a curved structure,
possibly a core plus jet (Murphy et al 1993).

\noindent{\bf MRC B1610$-$771}: Quasar with a flat radio spectrum and
very steep optical spectrum (Hunstead \& Murdoch 1980).  VLBI
observations (Preston et al 1989) show a strong core surrounded by a
50~mas halo.

\noindent{\bf MRC B1718$-$649}: The nearest GHz-peaked-spectrum
source, with a radio spectrum peaking near 3 GHz.  VLBI imaging shows
two sub-parsec-scale components separated by $\sim$2~pc (Tingay et
al.\ 1997).

\noindent{\bf MRC B1740$-$517}: Crowded optical field; galaxy ID by di
Serego Alighieri et al (1994) is confirmed by an AAT CCD image
(Burgess 1998).

\noindent{\bf MRC B1827$-$360}: Compact ultra-steep-spectrum source
identified with a galaxy in a very crowded field.

\noindent{\bf MRC B1829$-$718}: Candidate source for defining the VLBI
astrometric reference frame (Ma et al 1998).

\noindent{\bf MRC B1854$-$663}: Compact steep-spectrum source
identified with a faint galaxy (Burgess 1998).

\noindent{\bf MRC B1921$-$293}: See Paper I.

\noindent{\bf MRC B2052$-$474}: Radio spectrum steep at low frequency,
but flattens at high frequency; core dominated at 5 GHz, possibly
triple (Burgess 1998).  Detected as an X-ray source by the {\em ROSAT}\
All-Sky Survey (Brinkmann et al 1994).

\noindent{\bf MRC B2326$-$477}: Detected as an X-ray source in the
{\em ROSAT}\ All-Sky Survey (Brinkmann et al 1994).  One of the set of
defining sources for the VLBI astrometric reference frame (Ma et al 1997).

\subsection{General Properties}

If the observed variability is a result of refractive scintillation in
the Galactic interstellar medium (ISM), then we expect some sort of
dependence of one of modulation index $m=\sigma/\bar{S}$, characteristic timescale
$\tau_V$ or their product, $m\, \tau_V$, on the Galactic latitude, $b$
(e.g. Spangler et al. 1989; Ghosh \& Rao 1992).  However, apart from a
weak tendency for larger $\tau_V$ to occur at lower $|b|$,
there is no obvious correlation in our
data.  This is not surprising given the large uncertainties in
$\tau_V$ arising from the irregular sampling of the light curves, and
the fact that there are few variable sources at high latitudes (14 of
the 18 variable sources have $10^{\circ} < |b| < 30^{\circ}$).

\begin{figure}
\centerline{\psfig{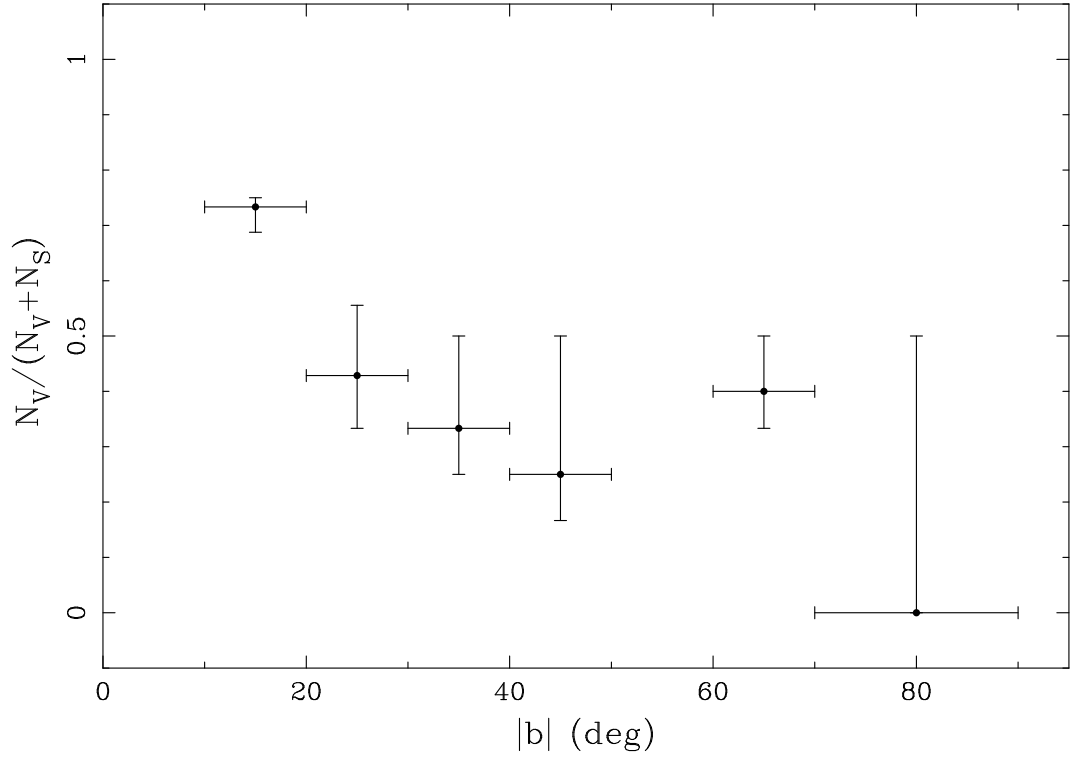}}
\caption{The fraction of variable sources as a function of Galactic latitude.
Horizontal error bars represent the width of each latitude bin, while 
vertical error bars have been derived by computing the fraction of variable
sources which results when half the undetermined
sources in that bin are reclassified as either variable (upper limits)
or steady (lower limits).}
\label{fig_hist}
\end{figure}

An alternative indicator of the effects of the Galactic ISM is to test
whether variable sources are more likely to be found at low
latitudes. We consider this possibility in Figure~\ref{fig_hist},
where we plot the ratio of variable sources ($N_V$) to variable plus
steady sources ($N_V + N_S$) in different latitude bins. While the
statistics are poor, there is a clear indication that sources are more
likely to be variable at low latitudes, as found for northern
hemisphere sources (Cawthorne \& Rickett 1985; Gregorini et al. 1986).
This is unlikely to be caused by selection effects associated with a
dependence of spectral index on Galactic latitude (cf. Cawthorne \&
Rickett 1985), since the main criterion for source selection was
angular size ($\theta < 10''$).  Thus the extensive monitoring data
for the MOST calibrators provide good evidence that the variability
observed at 843 MHz arises from scintillation in the local ISM.

While spectral index was not considered in selecting the majority of
the sample, Table \ref{tab_variables} shows that two-thirds of the variables
have flat or inverted spectra ($\alpha > -0.5$), consistent with
source angular size being the main determinant of source variability.
Surprisingly, the remaining third of the variables fall in the class of
compact steep-spectrum (CSS) sources which are generally believed to
be young sources still contained within their host galaxies, and not
known to vary at high frequencies.  The latter sources display a lower
level of variability, as measured by the modulation index $m$, and
in four of the six cases their V classification appears to be due to one-off
events lasting $\sim$1 year.

\begin{figure}
\centerline{\psfig{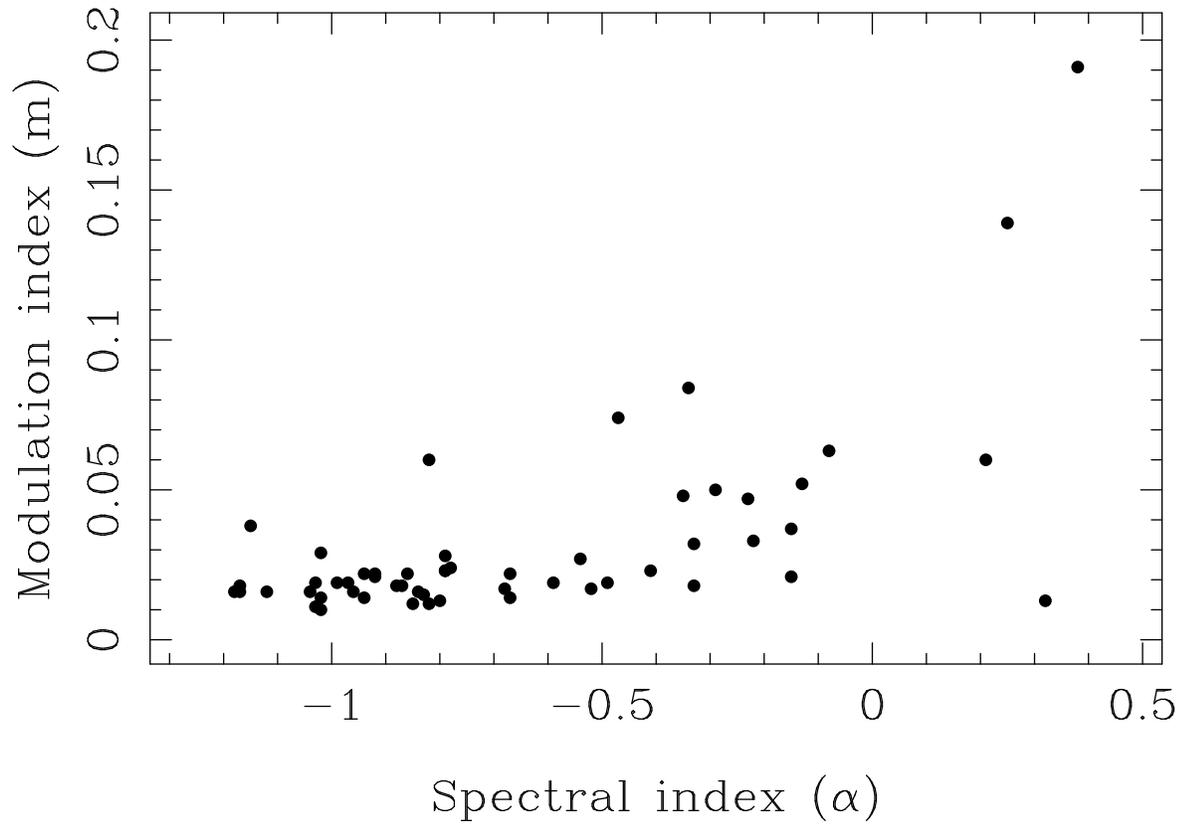}}
\caption{Plot of modulation index ($m$) versus spectral index
($\alpha$) for all 55 MOST calibrator sources.  Note the tendency for
$m$ to increase as the spectrum flattens.}
\label{m_alpha}
\end{figure}

To investigate the variability properties of the MOST calibrator
sample as a whole, in Figure~\ref{m_alpha} 
we have plotted $m$ versus $\alpha$ 
for all 55 sources.  This Figure shows a clear trend
towards higher average modulation index as the radio spectrum
flattens, with a suggestion of an upper envelope.  Perhaps the simplest
explanation for this behaviour in the unified model for
powerful extragalactic radio sources is to link $m$ and
$\alpha$ through the orientation of the radio axis to the line of
sight (e.g. Orr \& Browne 1982).  We assume that the `core' of a
classical triple source is the only part with components small enough
in angular size to scintillate.  If the core contribution dominates,
as a consequence of Doppler boosting in the flat spectrum sources,
even small fractional variations will be readily detected.  However,
the same fractional variations in the core of a steep-spectrum,
lobe-dominated source will go undetected.  We can therefore understand
the trends in Figure~\ref{m_alpha} in a qualitative sense, and it is
possible that a more detailed analysis may provide useful constraints
on radio source models.

\section{Conclusions}

55 sources used for calibration purposes by the
MOST at a frequency of 843~MHz have been
observed irregularly over a 13 year interval.
We have developed an algorithm to process these data and produce
a light curve for each source. Our analysis shows that 18 of these
sources can be considered variable. There is some suggestion that
these sources are distributed at lower Galactic latitudes than the 19
sources whose flux densities are unvarying. This suggests that
variability at 843~MHz on time scales of 1--10 years is predominantly
due to scintillation in the Galactic ISM rather than effects intrinsic
to the source.  A possible correlation between modulation index and
spectral index can be explained qualitatively in terms of a variation
in the core fraction with orientation of the radio axis to the line
of sight.

\small 

\section*{Acknowledgements}

We thank Ann Burgess for providing us with unpublished ATCA images of
several sources and for supplying us with her improved
meridian-distance gain curve.  We also thank Duncan Campbell-Wilson,
Lawrence Cram, Jean-Pierre Macquart,
Gordon Robertson, Mark Walker and Taisheng Ye for useful discussions
and advice, and an anonymous referee for a careful
reading of the manuscript.  
This research has made use of the NASA/IPAC Extragalactic
Database (NED), operated by JPL under contract with NASA.  The MOST is
supported by grants from the Australian Research Council, the
University of Sydney Research Grants Committee, and the Science
Foundation for Physics within the University of Sydney. BMG
acknowledges the support of an Australian Postgraduate Award and of
NASA through Hubble Fellowship grant HF-01107.01-98A awarded by the
Space Telescope Science Institute, which is operated by the Association
of Universities for Research in Astronomy, Inc., for NASA under
contract NAS 5--26555.


\section*{References}

\reference Bertsch, D.L. et al 1993 ApJ 405, 21

\reference Brinkmann, W., Siebert, J., \& Boller, T. 1994 A\&A 281, 355

\reference Burgess, A.M. 1998 PhD thesis, University of Sydney

\reference Campbell-Wilson, D., \& Hunstead, R.W. 1994 PASA 11, 33 (Paper I)

\reference Cawthorne, T.V., \& Rickett, B.J. 1985 Nature 315, 40

\reference di Serego Alighieri, S., Danziger, I.J., Morganti, R., \&
Tadhunter, C.N. 1994 MNRAS 269, 998

\reference Ghosh, T., \& Rao, A.P. 1992 A\&A 264, 203

\reference Gregorini, L., Ficarra, A., \& Padrielli, L. 1986 A\&A 168, 25

\reference Hughes, P.A., Aller, H.D., \& Aller, M.F. 1992 ApJ 396, 469

\reference Hunstead, R.W. 1972 Astrophys. Lett. 12, 193

\reference Hunstead, R.W. 1991 Aust J Phys 44, 743

\reference Hunstead, R.W. \& Murdoch, H.S. 1980 MNRAS 192, 31P

\reference Kaspi, V.M., \& Stinebring, D.R. 1992 ApJ, 392, 530

\reference Kedziora-Chudczer, L., Jauncey, D.L., Wieringa, M.H.,
Walker, M.A., Nicolson, G.D., Reynolds, J.E., \& Tzioumis, A.K.
1997 ApJ, 490, L9

\reference Kesteven, M.J.L., Bridle, A.H., \& Brandie, G.W. 1976 AJ 81, 919

\reference King, E.A. et al 1993, in `Sub-arcsecond Radio Astronomy',
ed. R.J. Davis \& R.S. Booth, Cambridge: CUP, 152

\reference Large, M.I., Mills, B.Y., Little, A.G., Crawford, D.F.,
\& Sutton, J.M. 1981 MNRAS 194, 
693\footnote{http://www.physics.usyd.edu.au/astrop/data/mrc.dat.gz}
 
\reference Large, M.I., Campbell-Wilson, D., Cram, L.E., Davison, R.G.,
\& Robertson, J.G. 1994 PASA 11, 44

\reference Ma, C. et al 1998 AJ 116, 516

\reference Mills, B.Y. 1981 PASA 4, 156

\reference Mitchell, K.J., Dennison, B., Condon, J.J., Altschuler,
  D.R. Payne, H.E., O'Dell, S.L., \& Broderick, J.J. 1994 ApJS 93,
  441

\reference Murphy, D.W. et al 1993, in `Sub-arcsecond Radio Astronomy',
ed. R.J. Davis \& R.S. Booth, Cambridge: CUP, 243

\reference Orr, M.J.L., \& Browne, I.W.A. 1982 MNRAS 200, 1067

\reference Preston, R.A. et al 1989 AJ 98, 1

\reference Qian, S.J., Britzen, S., Witzel, A., Krichbaum, T.P.,
Wegner, R., \& Waltman, E. 1995 A\&A 295, 47

\reference Rickett, B.J., Coles, W.A., \& Bourgois, G. 1984 A\&A 134, 390

\reference Robertson, J.G. 1991 Aust. J. Phys. 44, 729

\reference Shapirovskaya, N.Y. 1978 Sov. Astron. 22, 544

\reference Spangler, S., Fanti, R., Gregorini, L., \& Padrielli, L. 
1989 A\&A 209, 315

\reference Tingay, S.J. et al 1996 ApJ 464, 170

\reference Tingay, S.J. et al 1997 AJ 113, 2025

\reference Tzioumis, A.K. 1987 PhD thesis, University of Sydney

\clearpage

\begin{figure}
\centerline{\psfig{file=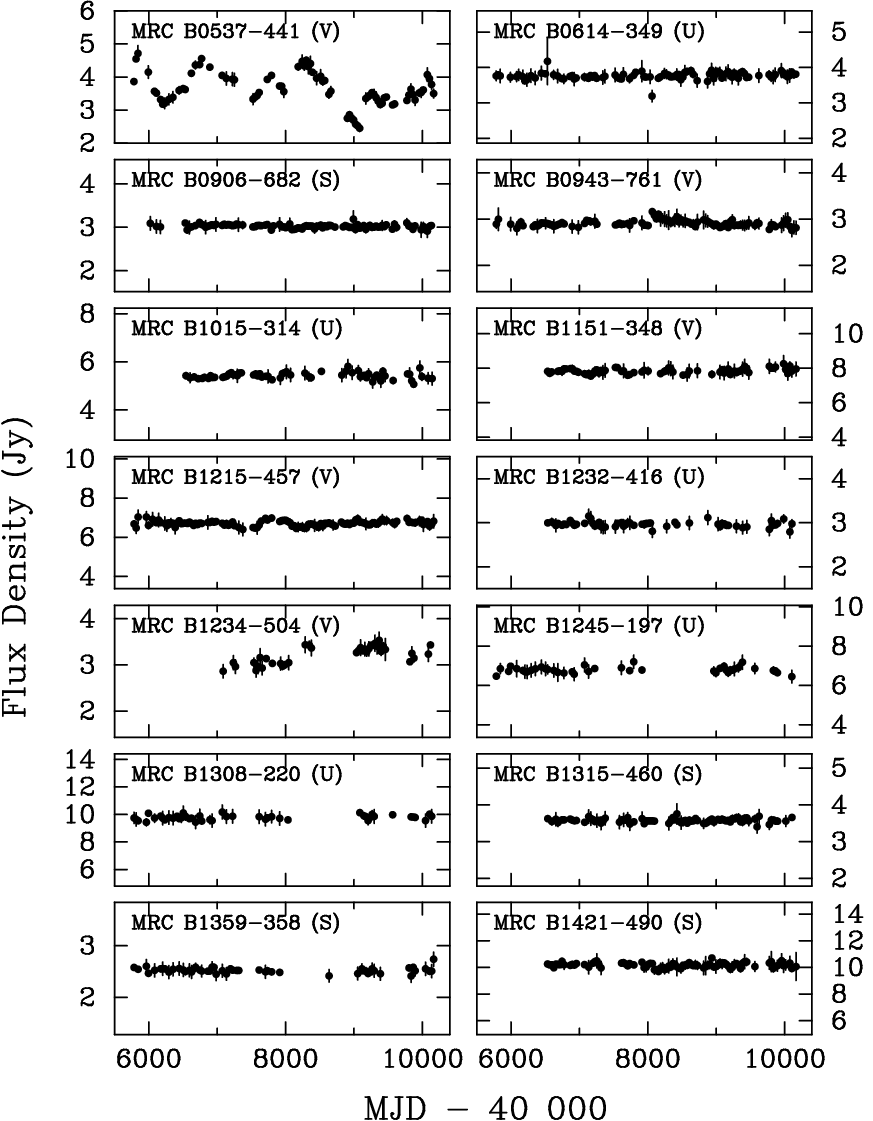,width=16cm,angle=270}}
Figure~\ref{fig_sources_1} (cont.)
\end{figure}

\begin{figure}
\centerline{\psfig{file=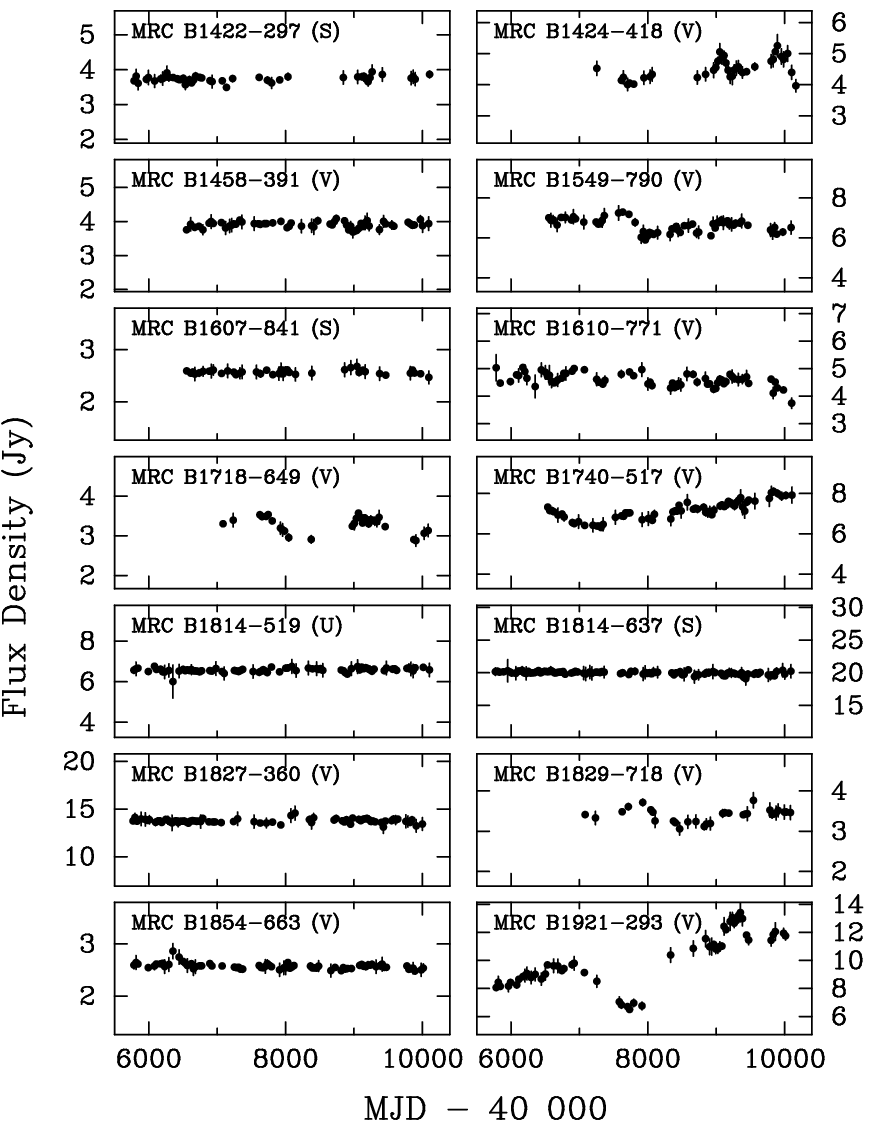,width=16cm,angle=270}}
Figure~\ref{fig_sources_1} (cont.)
\end{figure}

\begin{figure}
\centerline{\psfig{file=sources_4.ps,width=16cm,angle=270}}
Figure~\ref{fig_sources_1} (cont.)
\end{figure}

\begin{figure}
\centerline{\psfig{file=struc_2.ps,width=18cm,angle=270}}
Figure~\ref{fig_struc} (cont.)
\end{figure}

\clearpage

\begin{table}
\caption{Properties of variable sources in our sample.}
\label{tab_variables}
\begin{tabular}{lrrrclrr} \\[-3mm] \hline
Source & \multicolumn{1}{c}{$b$} & \multicolumn{1}{c}{$\tau_V$} & 
$m^a$  & Ident$^b$ & \multicolumn{1}{c}{$z$} & $\alpha^c$
 & \multicolumn{1}{c}{LAS$^d$}\\ 
 & (deg) & (days) & & & & & \multicolumn{1}{c}{($''$)} \\
\hline \\[-3mm]
MRC~B0208$-$512  & $-$61.8 &  2000 & 0.047 & Q & 1.003 & $-$0.23 & 6.0 \\
MRC~B0537$-$441  & $-$31.1 &  1500 & 0.139 & Q & 0.894 & +0.25 & \ldots \\
MRC~B0943$-$761  & $-$17.4 &  300  & 0.023 & g & \ldots & $-$0.79 & 2.8 \\
MRC~B1151$-$348  &  +26.3 &  400  & 0.019 & Q & 0.258 & $-$0.49 & $<$2.9 \\
MRC~B1215$-$457  &  +16.5 &  300  & 0.019 & Q & 0.529 & $-$0.59 & $<$1.9 \\
MRC~B1234$-$504  &  +12.0 & 1200  & 0.060 & Q? & $\ldots$ & $-$0.82 & $<$1 \\
MRC~B1424$-$418  &  +17.3 & 300   & 0.074 & Q & 1.52 & $-$0.47 & $<$2.3 \\
MRC~B1458$-$391  &  +17.0 & 400 &   0.022 & g & $\ldots$ & $-$0.67 & $<$2.4 \\
MRC~B1549$-$790  & $-$19.5 & 700 & 0.050 &   g & 0.15 & $-$0.29 & $<$1.0 \\
MRC~B1610$-$771  & $-$18.9 & 400 & 0.052 &   Q & 1.71 & $-$0.13 & $<$1 \\ 
MRC~B1718$-$649  & $-$15.8 & 1400 & 0.060 &  g & 0.013 & +0.21 & \ldots \\
MRC~B1740$-$517  & $-$11.5 & 2500 & 0.063 &  g & $\ldots$ & $-$0.08 & $<$1 \\
MRC~B1827$-$360  & $-$11.8 & 400 & 0.016 &  g & $\ldots$ & $-$1.12 & $<$1.5 \\
MRC~B1829$-$718  & $-$24.5 & 400 & 0.048 &  g & $\ldots$ & $-$0.35 & \ldots \\
MRC~B1854$-$663  & $-$25.5 & 400 & 0.022 &  g & \ldots & $-$0.86 & $<$1.0 \\
MRC~B1921$-$293  & $-$19.6 & 3000 & 0.191 & Q & 0.352 & +0.38 & \ldots \\
MRC~B2052$-$474  & $-$40.4 & 300 & 0.084 & Q & 1.489 & $-$0.34 & 3.9 \\
MRC~B2326$-$477  & $-$64.1 & 400 & 0.037 & Q & 1.489 & $-$0.15 & \ldots \\ 
\hline \\[-3mm]
\end{tabular}

$^a$ Modulation index, defined by $m = \sigma / \bar{S}$ \\
$^b$ Q = quasar ; g = galaxy\\
$^c$ Spectral index $\alpha$ (where $S\propto \nu^{\alpha}$) 
     between 408 and 2700 MHz (or 4850 MHz if 2700 MHz flux density not available)\\
$^d$ Largest angular size at 5 GHz, measured with the Australia Telescope 
     Compact Array (Burgess 1998)
\end{table}

\end{document}